\documentclass[doublecol]{epl2} 
\usepackage{amsmath,amssymb}
\usepackage{physics}
\usepackage{graphicx}
\usepackage{dcolumn}
\usepackage{bm}
\usepackage{xcolor}
\usepackage{subfigure}
\usepackage{amsfonts}
\usepackage{amssymb}
\usepackage{color}
\usepackage{cite}
\usepackage{hyperref}
\usepackage{float}

\title{Injection and   nucleation of topological defects in the quench dynamics of the Frenkel-Kontorova model}
\shorttitle{Injection and   nucleation of topological defects in the quench  
	dynamics of the Frenkel-Kontorova model} 

\author{Oksana Chelpanova\inst{1} \and Shane P. Kelly\inst{1} \and Giovanna Morigi\inst{2} \and Ferdinand Schmidt-Kaler\inst{1} \and Jamir Marino\inst{1}}
\shortauthor{O. Chelpanova, S. P. Kelly, G. Morigi, F. Schmidt-Kaler, J. Marino}

\institute{                    
	\inst{1} Institut für Physik, Johannes Gutenberg-Universität Mainz, D-55099 Mainz, Deutschland\\
	\inst{2} Theoretische Physik, Universität des Saarlandes, D-66123 Saarbrücken, Deutschland
}
\pacs{nn.mm.xx}{First pacs description}
\pacs{nn.mm.xx}{Second pacs description}
\pacs{nn.mm.xx}{Third pacs description}

\abstract{
	Topological defects have strong impact on  both  elastic and inelastic properties of  materials.
	In this article, we investigate the possibility to controllably inject topological defects in quantum simulators of solid state lattice structures. We investigate the quench dynamics of a Frenkel-Kontorova chain, which is used to model
	discommensurations of particles in cold atoms and trapped ionic crystals.
	The interplay between an external periodic potential and the inter-particle interaction makes lattice discommensurations, the topological defects of the model, energetically favorable and can tune a commensurate-incommensurate structural  transition.
	Our key finding is that a quench from the commensurate to incommensurate phase causes a controllable injection of topological defects at periodic time intervals.
	We  employ this mechanism to generate quantum states which are a superposition of lattice structures with and without  topological defects.
	We conclude by presenting concrete perspectives for the observation and control of topological defects  in trapped ion experiments.
}

\begin{document}
	
	\maketitle
	\section{Introduction} 
	The study of topological defects, such as solitons, is interdisciplinary, permeating  different areas of physics such as cold atoms~\cite{pitaevskii2016bose,PhysRevLett.106.130401,PhysRevLett.110.025302,PhysRevLett.104.160404}, spintronics~\cite{Magnetism,wieder2022topological,vedmedenko20202020}, and 
	nano-friction in  material science~\cite{liu2019recent,vanossi2020structural,Vuletic_nanofriction,PhysRevLett.115.233602,bohlein2012observation,PhysRevResearch.2.033198}.
	They are also at the core of a variety of technological applications encompassing quantum computing~\cite{tp1,tp2,tp3,topophotonics} and atomic sensors~\cite{atomtronics,bland2022persistent,eckel2014hysteresis}.
	In material science, they arise as lattice defects in solid state crystals and are particularly important in determining the mechanical properties such as rigidity~\cite{chaikin_lubensky_1995}.
	
	In this article, we investigate the possibility to controllably inject and eject topological defects in quantum simulators of solid state lattice structures, with the perspective of preforming similar nano-scale control of individual defects in real materials.
	To pursue this goal, we consider the quench dynamics of the Frenkel-Kontorova  model~\cite{frenkel1938see, FrankVanDerMerwe}, originally introduced to understand the macroscopic and static properties of dislocations in crystalline materials.
	This model can be simulated in a variety of quantum platforms such as ultra cold atoms~\cite{KasperMarino}, and trapped ion crystals~\cite{Ferdinand_exp,Drewsen_Dantan_2012,Vuletic_nanofriction,cetina2013one,Morigi3}; 
	unlike in materials, microscopic control over these systems is currently possible, allowing for experiments that can perform repeatable quenches which are sensitive to the quantum nature of the constituents particles.
	
	The topological defects in this model are lattice discommensurations induced by the interplay between an external periodic potential and  inter-particle interaction.
	By tuning  the external potential, the lattice can undergo a structural first-order transition from a (topologically trivial) commensurate state, in which particles stay at the  minima of an underlying periodic potential, to an incommensurate state, where  particles are dislocated from their natural equilibria, and the local discommensuration of particles can be mapped to a kink in a Frenkel-Kontorova model.
	
	Our main result is that these topological defects can be injected from the boundaries of the system  by performing quenches which originate from  the   commensurate  phase.
	This mechanism occurs in a controlled fashion with    atomic discommensurations   entering the chain  at  discrete intervals of  
	time   and building     a  metastable state characterized by a temporal staircase of  kinks, with life-time   scaling linearly with the chain size.
	We also explore   the impact of non-equilibrium fluctuations at the commensurate-incommensurate transition, and show that they can mold    novel states in the quantum Frenkel-Kontorova model. By quenching the system close to the   { boundary of the commensurate-incommensurate  transition,}
	quantum    fluctuations    lead to   {statistical} nucleation of  kinks, which drive the chain in a quantum superposition of  configurations with and without topological defects. 
	
	We focus throughout the paper on parameters regimes proper of  trapped ion simulators, and  we discuss in the concluding sections how to observe these phenomena in low dimensional ultracold atomic wires, arguing for broad applicability of our results to other quantum simulators. \\
	
	\section{Model}
	We consider $N$ atoms forming a one-dimensional chain with   inter-particle spacing $a_0 $. The  system is subject to   a periodic external lattice potential of period $a_s$. 
	The physics of the discommensuration between the  atoms spacing and the lattice periodicity can be captured  by 
	the Frenkel-Kontorova model~\cite{braun2004frenkel,Morigi3,Bak_1982,FrankVanDerMerwe,frenkel1938see}
	\begin{equation}\label{eq:model}
		H=\sum_n \frac{p_n^2}{2} +{m_K^2}(1+\cos\phi_n)+\sum_{r=\pm 1} \frac{(\phi_{n+r}-\phi_n-2 \pi r \delta )^2}{2} 
	\end{equation}
	\noindent where $p_n=\dot{\phi}_n.$
	Here the first term describes the kinetic energy of the atoms, the second one depicts the interaction with the lattice potential, and the last term captures the inter-particle interaction. Here  $\phi_n/(2\pi)=x_n/a_s-n$  denotes a classical field, characterizing the displacement of the $n$-th  particle position $x_n$ from the coordinate of the $n$-th minima of the lattice potential ($n=1,\ldots,N$).
	The dimensionless parameter $m_K^2$ captures the amplitude of the periodic lattice potential  and it plays the role of the kink mass in the following. 
	The `misfit parameter' $\delta=(a_0-a_s)/a_s$  quantifies instead the degree of discommensuration between $a_0$ and $a_s$.  
	
	\begin{figure}[]
		\includegraphics[width=\linewidth]{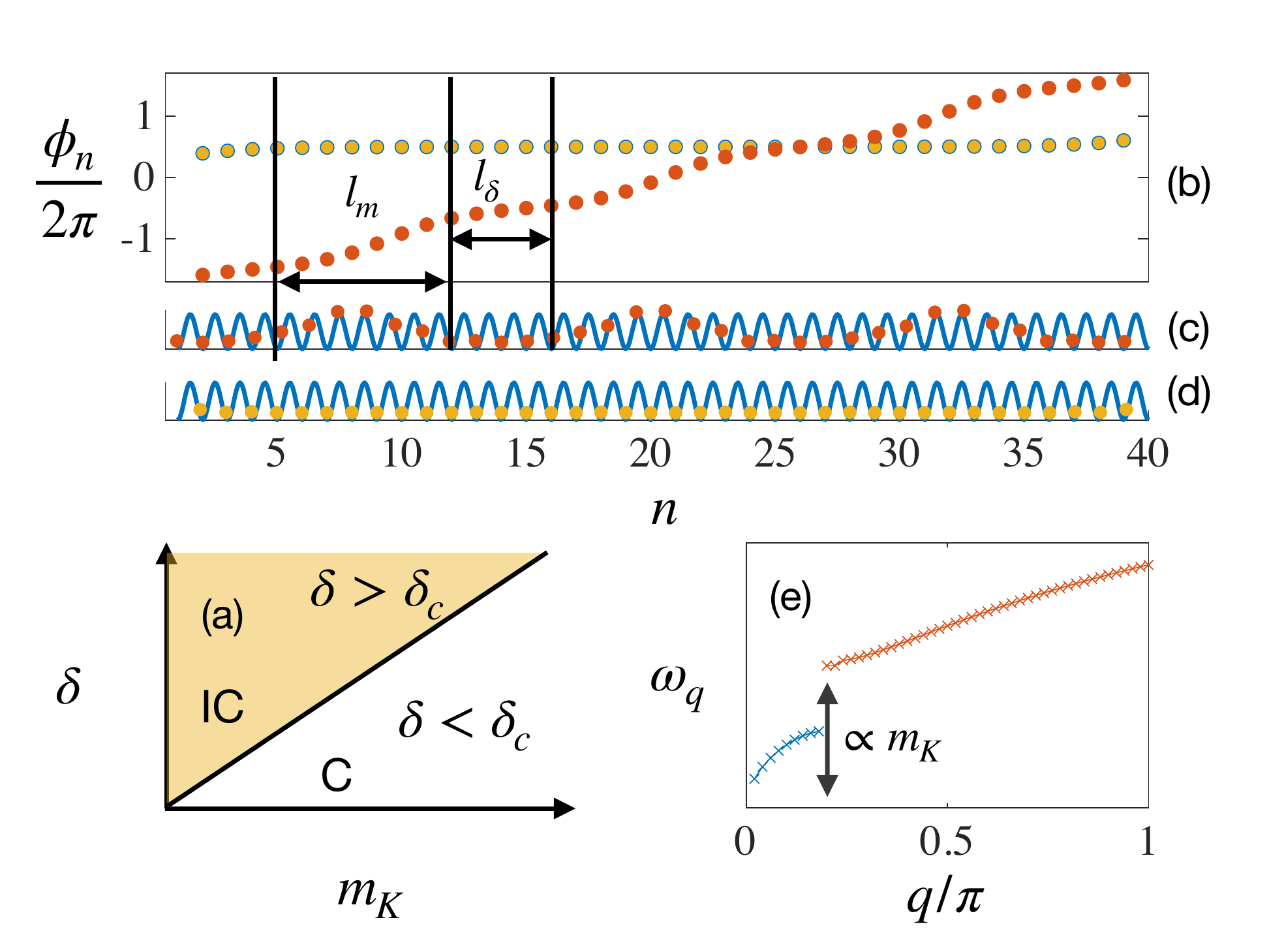}
		\caption{
			(a) Equilibrium phase diagram of the  model~\eqref{eq:model} as a function of the kink mass  $m_K$ and the misfit parameter $\delta$. 
			(b)  Displacement, $\phi_n$, of the $n$-th atom   with respect to the $n$-th  minimum of the potential  for 
			the incommensurate phase (orange) and the commensurate phase (yellow).  
			(c) Equilibrium positions of the atoms in the incommensurate phase. 
			Some atoms sit at the minima of the lattice potential, while others  in discommensurate equilibrium positions which can be   mapped into kinks of $\phi_n$. (d) The atomic positions  in the commensurate  phase coincide with the  minima of the trapping potential. (e) Sketch of the energy spectrum of the model as a function of momentum $q$ in the incommensurate phase. Blue crosses correspond to collective sonic excitations, while  red ones to parabolically dispersing modes (optical branch).}\label{fig:fig1}
	\end{figure}
	
	By tuning the amplitude and misfit parameter, $m_K$ and $\delta$, of the periodic lattice potential, the system at equilibrium undergoes a {structural} phase transition between two different configurations for $\phi_n$.
	The equilibrium phase diagram is sketched in Fig.~\ref{fig:fig1}(a). 
	For  $\delta<\delta_c\approx 2 m_K/\pi^2$ a strong  lattice potential pins the particles to the potential minima, forming a commensurate phase ($\phi_n=\pi$), see Fig.~\ref{fig:fig1}(b, d). 
	Instead, for values of $\delta\ge\delta_c$  atoms reside in  a new incommensurate configuration  associated   with kinks in $\phi_n$ placed at the location of the   particles' displacements, cf. Fig.~\ref{fig:fig1}(b, c).  {
		The kinks (anti-kinks) correspond to a local distribution of holes~(excess particles) in the chain~\cite{Morigi3,Bak_1982,braun2004frenkel,Pruttivarasin_2011,garcia2007frenkel,Bak_1982}}, and are represented by a    `step' in the displacement field, $\phi_n$, equal to $2\pi$ ($-2\pi$). 
	The number of  kinks $Q=(\phi_N-\phi_1)/(2\pi)$ is {a parameter} that allows us to distinguish between the commensurate and the incommensurate phases~\cite{AristovLuther}.
	This quantity estimates the number of `steps' in $\phi_n$,  {i.e., jumps in the value of the field $\phi$ from $\pi+2\pi m$ to $\pi+2\pi(m+1),$ where $ m$ is an integer number}. For instance, the incommensurate configuration in Fig.~\ref{fig:fig1}(b, c) has three `steps' in $\phi_n$ which corresponds to $Q=3.$ 
	For large values of the misfit parameter $\delta$,   the number of kinks, $ Q$, is linearly proportional to $\delta$.  {	Accordingly, $Q$ is directly related to  
		the spacing between kinks, $l_{\delta}\propto 1/\delta.$ 
		
		In the following we investigate the dynamical injection of kicks and use the width of the kink and the excitation spectrum as key parameters to interpret.
		The width of the kink, $l_m$, is   equal to the distance over which  $\phi$ increases by $2\pi$, and it is roughly proportional to  the inverse 
		of the kink mass, 
		$l_m\propto 1/m_K$ \cite{KasperMarino,AristovLuther,Morigi3}. For large $m_K$ the shape of the kink is sharp, since   any broadening of the incommensurate region  requires additional energy cost due to the lattice potential.
		For small $m_K$ the lattice potential is almost negligible and even a single kink appears wide.}
	Following~\cite{AristovLuther, EntanglementStatic}
	the
	excitation spectrum of the model~\eqref{eq:model} in the two phases    {can be extracted}  by expanding the equations of motion for $\phi_n$ around the classical solution $\phi_n^{(0)}$ in powers of small fluctuations  {$\eta_n$}. The corresponding equations of motion for the linear displacements, $\eta_q$, in momentum space read 
	\begin{equation}\label{eq:spectrum}
		\omega_{q}^{2}\eta_{q}=-{m_{K}^{2}}\cos(\phi^{0})\eta_{q}+c^2 q^2 \eta_q+\ldots,
	\end{equation}
	\noindent  {where $c$ is the speed of sound, which is equal to one for the model~\eqref{eq:model}}.
	The  spectrum  in the incommensurate phase consists of  two branches, cf.~Fig.~\ref{fig:fig1}(e). The first  $Q$ modes correspond to gapless collective excitations,  phonons, describing  the propagation of kinks through the lattice at the speed of sound $c$ (blue line).   The remaining $N-Q$ modes form an  optical branch (red line) and they describe small harmonic displacements of the atoms around the potential minima: they are parabolically dispersing excitations with a gap set by $m_K$. Within the commensurate phase, all the $N$ modes are optical.

	\section{Controlled Injection of Defects} 
	We now show how to inject, and also eject, topological defects by keeping $m_K$ fixed, and making an abrupt change of $\delta$, from $\delta_i$ to $\delta_f,$ which in terms of experimental parameters  corresponds to the sudden change of the inter-particle spacing $a_0$.
	We start by solving numerically the equations of motion that govern the dynamics of the topological defects in model~\eqref{eq:model}, which read 
	\begin{equation}\label{eq:eom}
		\begin{aligned}
			&	\ddot{\phi}_1  = m_K^2\sin \phi_1+\left(\phi_{2}-\phi_1 -{2\pi } \delta\right) \\
			&		\ddot{\phi}_{n=2:N-2} ={m_K^2}\sin \phi_n+\\
			& +\left(\phi_{n+1}-\phi_n -{2\pi } \delta\right)+\left(\phi_{n-1}-\phi_n +{2\pi } \delta\right)\\
			&		\ddot{\phi}_N  ={m_K^2}\sin \phi_N+\left(\phi_{N-1}-\phi_N +{2\pi } \delta\right).
		\end{aligned}
	\end{equation}
	Here the effect of the discommensuration between $a_0$ and $a_s$ cancels for particles in the bulk, as it produces opposite contributions from interaction with right and left neighbours. However, 
	the misfit parameter affects directly dynamics of the boundary particles, because the contribution from discommensuration is not compensated.
	In this way, after the quench of $\delta$ only the edge is initially driven out of equilibrium such that the boundary acts as a reservoir for topological defects (see also~\cite{PhysRevA.91.033607,PhysRevLett.110.025302,PhysRevE.100.062202,ma2020realization}).
	Furthermore, due to their topological nature, once the defects enter the bulk, they cannot be destroyed~\cite{etde_6020750,dauxois2006physics,Brox_kinks_spectroscopy}.

\begin{figure}[]
		\includegraphics[width=\linewidth]{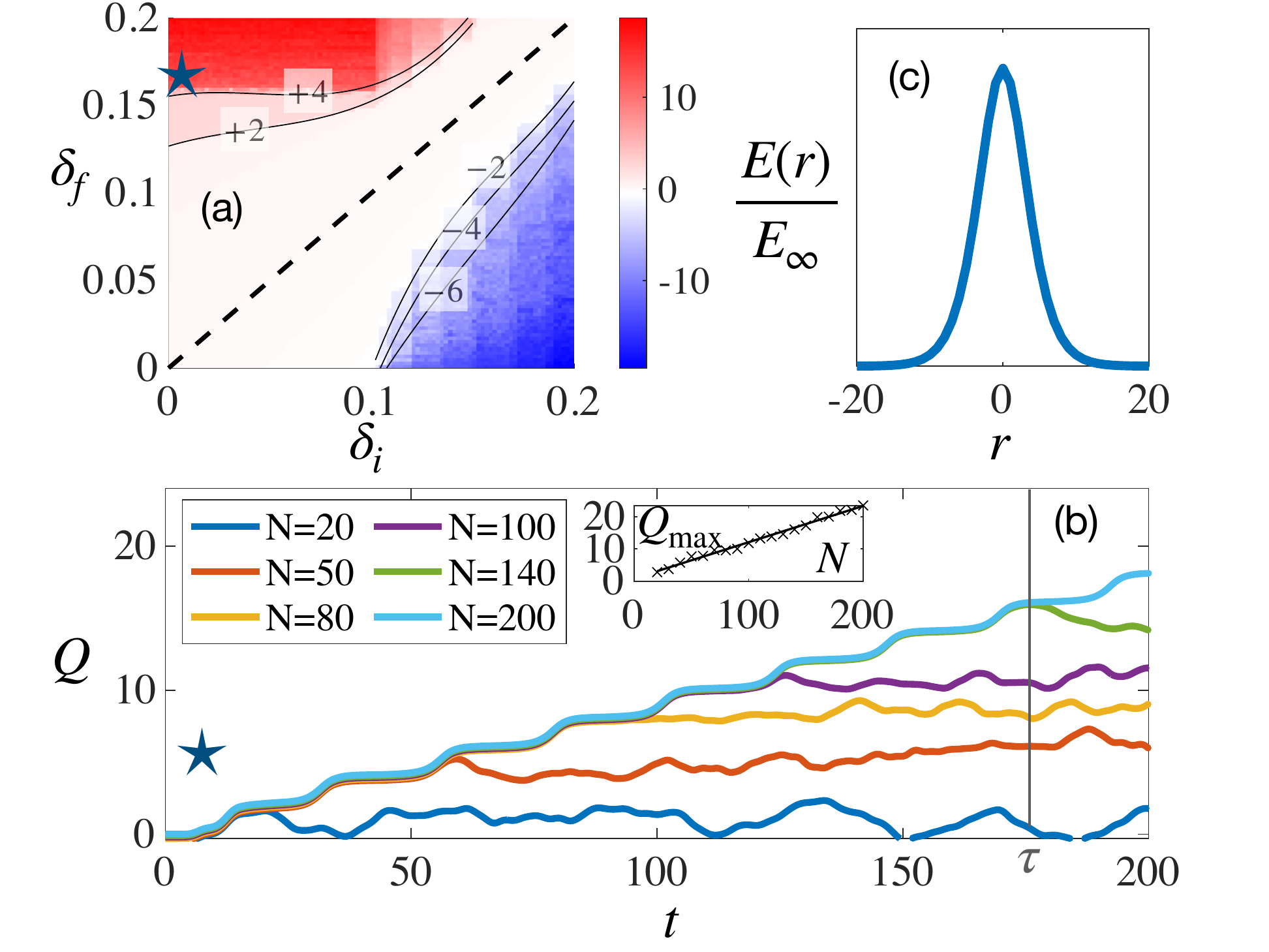}
		\caption{   
			(a) Dynamical response  of the system after a quench of the  misfit parameter from $\delta_i$ to $\delta_f$,   with $m_K=0.5$ (here $N=80$). We plot  the time-averaged   number of kinks, $ \bar{Q}-Q_{(t=0)}$. 
			{The solid lines mark the critical thresholds,  $\delta_c^d $,  corresponding to the discrete injection of  a certain   number of (anti-)kinks.}
			The dashed line corresponds to  $\delta_i=\delta_f$. The red region corresponds to the injection of  kinks into the system, while the blue one pertains to anti-kinks. 
			(b) Dynamics of the number of kinks in the system after the quench marked with a $\color{blue}\star$ in  panel (a). Here we fix $\delta_i=0$ and $\delta_f=0.16>\delta_c^d$ and consider different system sizes $N$ (marked with different  colors). The plot illustrates periodic injection of kinks mirroring in a `step-like' increase of $Q$ in time. 
			The time scale $\tau\propto N/c$ denotes  the saturation time of $Q$ for $N=140$ {ion}s after which finite-size effects  take over.
			The maximal number of kinks in the system after the quench scales linearly with $N$ (see inset). 
			(c)  {Energy of  two kinks as a function of the  relative distance between   their centers. The effective kinks repulsion  takes place at distances   $r< l_{\delta}+l_m$. The energy scale in panel (c) is normalized by the energy of  two infinitely separated kinks $E_{\infty}$.}
		}\label{fig:fig1_2}
	\end{figure}

	In order to see this concretely, we numerically initialize a  system in the equilibrium state  with misfit parameter $\delta_i$, and then numerically solve its evolution according to equation~\eqref{eq:eom} with the misfit parameter $\delta_f$. 
	In Fig.~\ref{fig:fig1_2}(a) we plot a diagram  of the possible dynamical responses in the system using the long-time averaged number of kinks  $\bar{Q}$, where time  averages are taken over several periods $\propto 2\pi/m_K$.
	In the white region of Fig.~\ref{fig:fig1_2}(a) no kinks or anti-kinks are injected during dynamics. 
	The red region in the phase diagram corresponds to      kinks' injection  after
	quenching  the misfit parameter above a  dynamical critical threshold,  $\delta_c^d$. Finally,    
	the blue region in Fig.~\ref{fig:fig1_2}(a) corresponds to quenches which decrease  the  total number of kinks. This occurs by  injecting anti-kinks into the system. 
	Injection of a {single} kink (an anti-kink) from the boundary is  associated with the excitation of a collective sonic mode.

	Fig.~\ref{fig:fig1_2}(b) shows
	how the
	number of kinks in the system increases
	following the quench from the commensurate to a target point of the incommensurate phase, marked with $\color{blue}\star$  in   Fig.~\ref{fig:fig1_2}(a). Dynamics start from $Q=0$ and exhibit a staircase structure in time due to the injection of kinks from the boundaries up to time $\tau\propto N/c$ when  
	finite-size
	effects become relevant. 
	The duration  of the various plateaus of the staircase is proportional to $l_{\delta_f}/c$. Arranging kinks at $r<l_{\delta}+l_m$ requires an additional energy cost. In order to show that, we calculate the energy of  two kinks as a function of the relative distance between their centers $r$, cf. Fig.~\ref{fig:fig1_2}(c). If the kinks overlap, their energy  is higher than the energy of the configuration  with two infinitely separated kinks,  resulting in short-distance    repulsion~\cite{CARRETEROGONZALEZ2022106123,braun2004frenkel}. 
	Thus,  
	the average distance between kinks after the quench is fixed  to approximately  $ l_m+l_{\delta_f}$. As a result, the  number of kinks saturates at long times to $Q\approx  2 N/(l_m+l_{\delta_f})$, which is basically given by the total number of solitons fitting in the system, for a given size $N$  (cf. inset in Fig.~\ref{fig:fig1_2}(b)). Note, that $\propto N/(l_m+l_{\delta_f})$ kinks enter the system from each boundary.
	Indeed, up to times $\tau$, systems with a different number of particles evolve   with the same staircase temporal profile, since dynamics are  universally ruled by ${\delta_f}$ and $m_K$.   
	Notice that the height of each `step' in Fig.~\ref{fig:fig1_2}(b) is equal to two,  since we simultaneously inject one kink from the left   and another one from the right boundary, after the quench. 
	Finally, at late  times $t>\tau$, the number of kinks, $Q(t)$, oscillates around the steady state value, see  Fig.~\ref{fig:fig1_2}. This behaviour  results from both sonic and optical modes. The first one alters the value of charge as a result of  scattering of the injected kinks against the boundaries. The second one describes individual oscillations of particles around potential minima.

	\section{Dynamics close to the  {boundary of the commensurate-incommensurate  transition}}  An exciting feature of cold atom and trapped ion platforms is their sensitivity to quantum fluctuations. 
	Generally, the effect of quantum fluctuations within a given phase only quantitatively modify the dynamics, but the dynamics close to 
	{the boundary between two phases} can have non-trivial effects such as destroying order, generating  new dynamical phases, or provoking  a  quantum critical region  in the system  \cite{sachdev,vestigial_order,Giovanna_zigzag_PRL, nucleation,Giovanna_zigzag_PRA,Lazarides_finite_T_RG,AlessioPRL,AlessioPRB,PhysRevLett.123.230604,Marino_2022}. 
	As such, we investigate the effect of quantum fluctuations for quenches close to the boundary of the commensurate-incommensurate transition, and  we introduce the quantum version of Hamiltonian~\eqref{eq:model}:
	\begin{equation}\label{eq:qm_model}
		\hat{H}=\sum_n\frac{\hat{p}_n^2}{2} +m_K^2(1+\cos\hat{\phi}_n)+\sum_{r=\pm 1} \frac{(\hat{\phi}_{n+r}-\hat{\phi}_n-2 \pi r \delta )^2}{2} 
	\end{equation}
	where we have promoted $p_n$ and $\phi_n$ to the dimensionless quantum operators $\hat{p}_n$ and $\hat{\phi}_n$ with canonical commutation relations $[\hat{\phi}_n,\hat{p}_m]=ih_{\operatorname{eff}}\delta_{n,m}$.
	Here we take $h_{\operatorname{eff}}$ as a dimensionless free parameter which controls the strength of quantum fluctuations. 
	Below we give an expression for $h_{\operatorname{eff}}$ for the trapped ion implementation. 
	
	In the following, we take $h_{\operatorname{eff}}\ll 1$ and investigate the impact of  quantum fluctuations using the semi-classical approximation known as the Truncated Wigner Approximation (TWA)~\cite{POLKOVNIKOVTWA,PNAS_TWA,PhysRevA.100.013613}. 
	We consider a quench from the ground state of the Hamiltonian~\eqref{eq:qm_model},  that can be approximated~\cite{POLKOVNIKOVTWA} by a Gaussian wave function with the Wigner function equal to
	\begin{equation}\label{eq:TWA}
		W\left(\eta_{q}^{0},n_{q}^{0}\right)=\prod_{q}\exp\left[-{\left|\eta_{q}^{0}\right|^{2}}/{\sigma_{q}}-{\left|\psi_{q}^{0}\right|^{2}}/b_{q}\right],
	\end{equation}
	\noindent where $\eta^0_q=\phi^0_q-\langle \hat{\phi}_q\rangle_0$ and $\psi^0_q=p^0_q-\left<\hat{p}_q\right>_0$ are the phase space variables relative to the mean value of the operators in the initial state,  while $\sigma_q=\langle\hat{\eta}_q^2\rangle_0=h_{\operatorname{eff}}/\omega_q$ and $b_q=\langle\hat{\psi}_q^2\rangle_0=h_{\operatorname{eff}}\omega_{q}$ encode the variance  of quantum fluctuations in the initial state. Finally, the dispersion relation $\omega_q$ is determined by Eq.~\eqref{eq:spectrum}.
	At zeroth order in $h_{\operatorname{eff}}$, the TWA approximation is simply the mean field dynamics discussed above where the system evolves by the classical equations of motion Eq.~\eqref{eq:eom} with initial state taken as the mean field ground state.  
	At first order in $h_{\operatorname{eff}}$, the dynamics of phase space variables are still described by Eq.~\ref{eq:eom}, but the initial state is sampled from the positive Wigner distribution in Eq.~\eqref{eq:TWA}. 
	Observables at later times are calculated by averaging over the trajectories resulting from the initial quantum noise. 
	Taking into account higher orders quantum corrections can be performed by including quantum jumps during dynamics in each trajectory, with each  jump introducing  $O\left(h_{\operatorname{eff}}^2\right)$ corrections~\cite{POLKOVNIKOVTWA,TWA1stcorr}.
	
	\begin{figure}[]
		\centering
		\includegraphics[width=\linewidth]{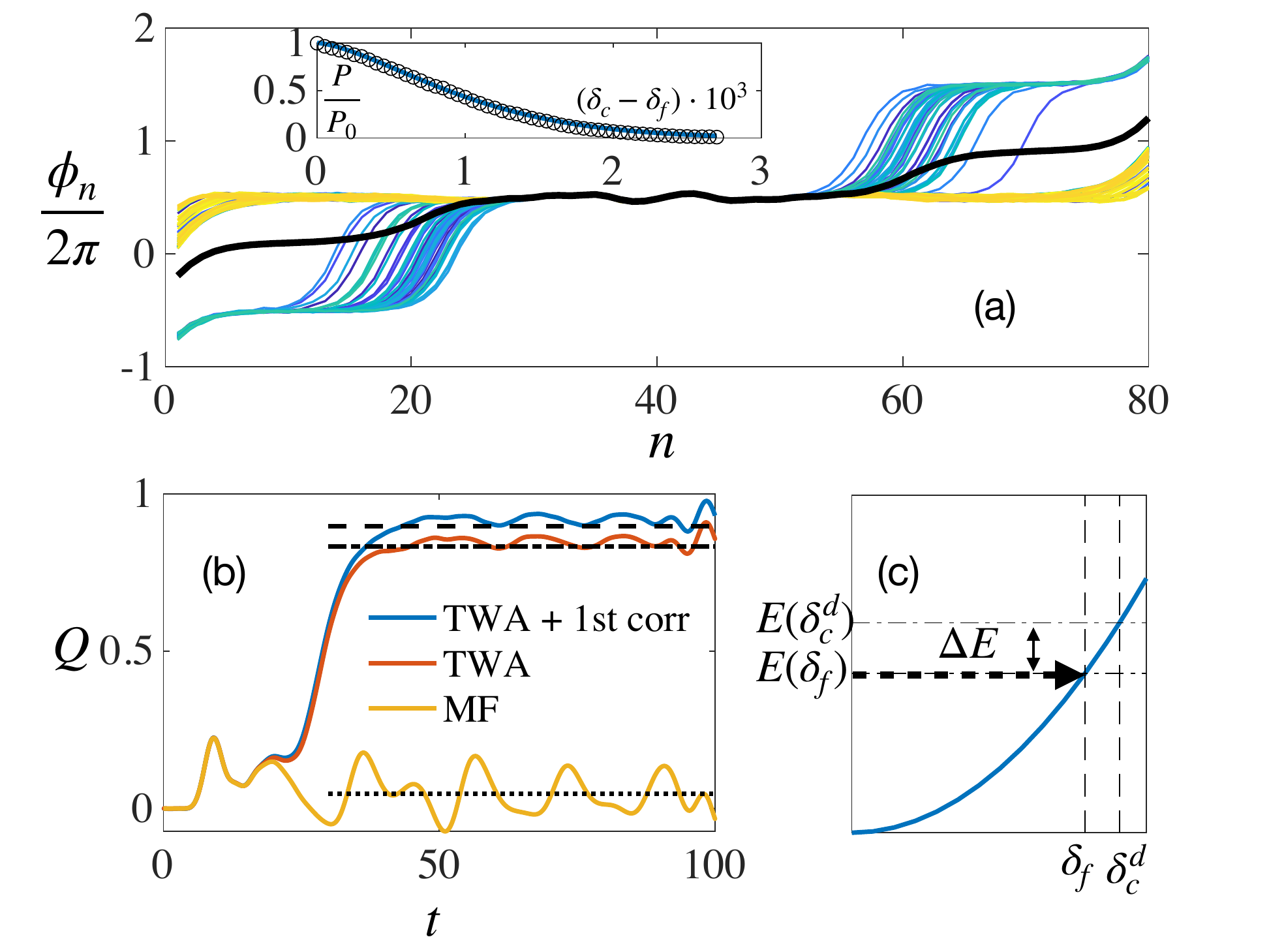}
		\caption{
			(a) Separation of trajectories after a quench close to the  {boundary between two phases $\delta_i=0\to\delta_f\lesssim\delta_c^d$ }.
			Figure shows different particle displacements, $\phi_n$,  following the quench at a time $t\propto N/c$. Different configurations correspond to different initial conditions sampled from the distribution in Eq.~\eqref{eq:TWA}, and color distinguishes between configurations that remain in the commensurate phase (yellow) and those that nucleated a kink (blue). 
			The overall number of kinks in the averaged value of  $\phi_n$  (solid black line), at long times,  is non-vanishing. (b)  Number of kinks as a function of time after the quench calculated classically (yellow), semi-classically (red), and taking into account  next-order  quantum corrections (blue).
			Dashed lines correspond to time-averaged $Q$. The  number of kinks at the classical level is zero and the system remains in the commensurate phase, while for TWA dynamics  and for its first-order correction, the time-averaged  $Q$ takes a non-integer value between zero and one.
			(c) Energy of the system  as a function of $\delta$ calculated numerically. For quenches $\delta_f<\delta_c^d$ entering the incommensurate phase after the quench requires  additional energy cost $\Delta E=E(\delta_c^d)-E(\delta_f)$, which makes this transition impossible in the classical case. Taking into account quantum fluctuations allows system to tunnel this energy barrier and nucleate a kink. 
			The  inset of panel (a) compares the probability of nucleating a kink through the energy barrier as a function of the final misfit parameter, for TWA numerics (circles) and for the WKB approximation (solid); fitting parameters are $P_0=0.81$ and $\alpha=5.$
		}
		\label{fig:fig2}
	\end{figure} 
	
	Let us now revisit the quench of the misfit parameter within the commensurate phase, $\delta_i=0\to \delta_f\lesssim \delta_c^d.$
	At the classical level this excites the displacement 
	field  $\phi_n^{(0)}$ in an optical mode. At small times the dynamics of  fluctuations are  governed by  $\Ddot{\eta}_q\approx m_K^2 \cos \left(\phi^0(t)\right) \eta_q-c^2 q^2 \eta_q$ (cf. with~Eq.~\eqref{eq:spectrum}), with    a periodically driven mass term set by the motion of $\phi_n^{(0)}(t)$. 
	In cosmology~\cite{Starobinsky} or cold atoms~\cite{PolkovnikovQBreathers}, a similar structure for   dynamics of the   low-energy modes results in an instability that can trigger  non-linear effects. In our   system 
	the parametric amplification of low-energy modes can indeed cause the  separation of quantum trajectories on times of order $\log{h_{\operatorname{eff}}}$.
	Fig.~\ref{fig:fig2}(a) shows
	final configurations of the field for different samplings of the  Wigner distribution associated to the pre-quench state.
	The yellow trajectories remain within the commensurate phase and contain no kinks, while the blue trajectories are the 
	ones which have tunnelled into the incommensurate region, and in fact they display kinks in the profile of the
	displacement field, $\phi_n$. 
	Such a semi-classical ensemble represents a quantum state that is a superposition of an atomic lattice in the commensurate configuration and a configuration with a kink.
	We check that this result remains qualitatively unaltered 
	after  adding higher order quantum corrections in Fig.~\ref{fig:fig2}(b).
	
	Classically, a quench  below the dynamical critical threshold, $\delta_i=0\to \delta_f\lesssim \delta_c^d $, keeps the system within the commensurate phase since the injection of a single kink  would require an additional energy cost
	$\Delta E=E(\delta_c^d)-E(\delta_f)$. 
	By adding quantum fluctuations, however, the system can tunnel through this barrier and nucleate a kink at the boundary. 
	The fraction of trajectories that nucleated a kink after the quench of $\delta$ can be estimated via WKB approximation~\cite{berry1972semiclassical}. In this approximation, nucleation of a kink 
	can be understood as tunneling of a quantum particle through the energy barrier separating a commensurate from an incommensurate configuration, see Fig.~\ref{fig:fig2}(c).
	The transition probability according to WKB reads
	\begin{equation}\label{eq:tunneling}
		P=P_0\exp\left( -{\alpha}{h_{\operatorname{eff}}^{-1}}\int_{\delta_f}^{\delta_c^d}\sqrt{E(\delta_c^d)-E(\delta)}d\delta\right),
	\end{equation}
	which we fit   in  good agreement with numerical results in the inset of Fig.~\ref{fig:fig2}(a).
	Here the effective Planck constant $h_{\operatorname{eff}}$ determines  the width of the region in $\delta_c^d-\delta_f$ where nucleation of kinks can be observed, namely, $\sqrt{\Delta E}(\delta_c^d-\delta_f)\lesssim h_{\operatorname{eff}}\alpha^{-1}$.
	Quenches away from this region do not produce states with the above quantum superposition. Instead, quantum noise provokes dephasing  among trajectories and can result in the broadening of the width of a single kink such that atoms will be harder to resolve. 
	
	Similar arguments hold for quenches deep into incommensurate region.  
	As shown in the portrait of  dynamical responses  (Fig.~\ref{fig:fig1_2}(a)) there is a set of critical thresholds $\delta_{c_{n}}^d$ that correspond to the injection of multiple ($n=\pm 2,\pm 4,\ldots$) solitons into the system. Therefore, by quenching $\delta$ in between $\delta_{c_{n}}^d$ and $\delta_{c_{n+2}}^d$ we can generate
	states which are in a superposition of
	$n$ and $n+1$ kinks.

	Notice that if the initial state was instead a Gibbs thermal state, transitions into the incommensurate state could also occur by thermal activation~\cite{reichl1999modern}, and it would give rise  qualitatively  to the same results of Fig.~\ref{fig:fig2}.
	The major difference is that the system would be in a mixed state of crystal configurations with different values of $Q$ and with quantum coherence decreasing with increasing temperature.
	
	\begin{figure}[]
		\centering
		\includegraphics[width=\linewidth]
		{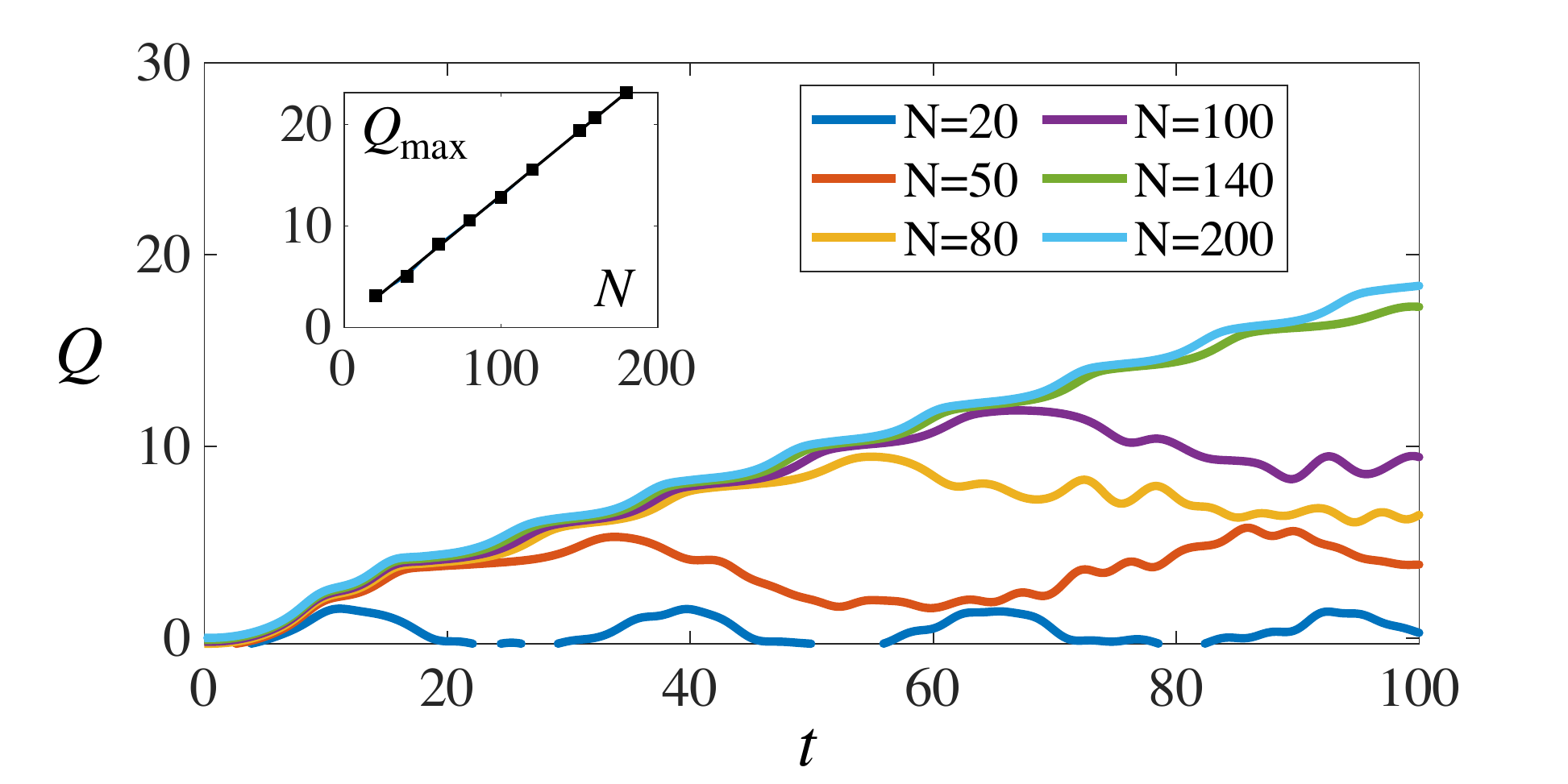}
		\caption{Number of kinks in a chain  after the quench of the misfit parameter from the commensurate to the incommensurate phase in a long-range interacting model (cf. Eq.~\eqref{eq:Hlongrange}) for different system sizes $N.$ Here we fix $m_K=0.5$ and $\delta_f=0.11. $
			Inset shows the scaling of the maximal number of kinks, injected after the quench as a function of the system size. 
		} \label{fig:fig1_SM}
	\end{figure}
	
	\section{Trapped Ion Realization\label{sec:trapped_ion}}
	We now discuss how the quench dynamics presented in the core of the paper can find application in trapped ion chain~\cite{Ferdinand_exp,Drewsen_Dantan_2012,Vuletic_nanofriction,cetina2013one}.
	An ionic crystal is formed by trapping ions in a 1D Paul trap whose strength sets the inter-particle distance $a_0$.
	The external lattice potential is provided by an optical lattice with period $a_s$, and amplitude, {$\epsilon$, that sets the mass $m_K^2=(2\pi)^{2}{\left(4\pi\epsilon_{0}\right)\hbar\epsilon a_{0}^{3}}/{\left(4q^{2}a_{s}^{2}\right)}$.} Here $q$ is an electron charge and $\epsilon_{0}$ is a vacuum permittivity.
	The effective Plank constant takes the form $\hbar_{\operatorname{eff}}={\hbar(2\pi)^{2}}\sqrt{{(4\pi\epsilon_{0})a_{0}^{3}}/({2q^{2}M})}/{a_{s}^{2}}, $ where $M$ is the mass of a single $^{40}$Ca ion. 
	In a trapped ion implementation the parameters of model~\eqref{eq:model} can be changed in the range $0<m_K<1$, $-0.5\le \delta\le 0.5$ and $h_{\operatorname{eff}}\le 10^{-2}.$
	The discussion above holds except for an unscreened Coulomb interaction that creates a long range modification to the harmonic interaction between neighboring ions.
	In this section we explore how the long range interaction modifies the ability to controllably inject topological defects into the ionic lattice by considering the dynamics under the modified Hamiltonian
	\begin{equation}\label{eq:Hlongrange}
		{H}= \sum_n  \frac{p_n^2}{2}  +
		m_K^2 ( 1+\cos\phi_n ) +
		\sum_{r>0}\frac{(\phi_{n+r}-\phi_{n} -2\pi r \delta )^2}{2  |r|^3} 
	\end{equation}
	where $r$ is an integer and runs over all neighbours.

	As the inter-particle interaction increases, the form of a single kink modifies and  it becomes less sharp  on the tails~\cite{Morigi3,PhysRevB.41.7118}. As a result, the effective interaction between kinks is also modified.
	Furthermore,   in case of the long-range interaction all ions are   sensitive to the presence of dislocations, and thus, injection of each kink into the system contributes immediately into dynamics of the ions in the middle of the chain.  
	Nonetheless, as we show below,  our previous results  remain qualitatively similar.

	Similarly to the model with short range interactions, quenching  the misfit parameter can drive commensurate-incommensurate transition in the long-range interacting model~\eqref{eq:Hlongrange}. Fig.~\ref{fig:fig1_SM} shows the number of kinks in the system $Q$ as a function of time in this case, and again shows kinks entering chain one by one from the boundaries in a periodic time intervals. 
	Quantitatively, the duration of the plateau is shortened as the form of a single kink is broadened by  the Coulomb repulsion. 
	Furthermore, in the short-range interacting model injection of kinks takes place independently  from left and right boundaries and it continues up to the time when first injected kink reaches the opposite boundary. 
	However, in the long-range interacting model, dynamics of all atoms in the chain is immediately affected by the presence of the topological defects. Thus,
	the saturation time for a long-range interacting system, is approximately two times shorter than for its short-range interacting counterpart, $\tau\approx N/2 c.$ It results from the fact that at this time $N/2(l_m+l_{\delta})$ kinks entered the chain from right boundary and  $N/2(l_m+l_{\delta})$ kinks entered the chain from the left and the given density of kinks already minimizes the total energy of the system.

	The role of Coulomb repulsion can be effectively suppressed  by increasing $m_K$, whose inverse sets its relative strength with respect to the lattice potential.  In this case, a staircase structure in $Q(t)$ can be observed more clearly.  
	Even in a long-range interacting model the maximal number of kinks in the system scales linearly with the system size, sf. inset in Fig.~\ref{fig:fig1_SM}, which is also the case for the nearest-neighbour interactions.

	In order to observe these dynamics in experiments, it is necessary to resolve the commensurate and incommensurate  configurations.
	One possibility would be to image the    ions positions  by applying non-destructive far-field measurements~\cite{Schmidt_Kaler_far_field} using the diffractive  pattern obtained from the scattered   light.
	Alternatively, one could resort to AC-Stark shift methods~\cite{Schmidt_Kaler_ac_Stark} in order  to distinguish the positions of the   ions more accurately. 
	The above results pertain to relevant  parameters regime of a typical trapped ion experiment~\cite{Ferdinand_exp,Vuletic_nanofriction}, and   experimental details such as the effect of  the external Paul trap, and an estimate  of the temperature required 
	to observe quantum suppositions of kinks in trapped ions experiments 
	are elaborated in~\cite{solitons_long}.
	
	\section{Outlook} 
	The effective long-wavelength description of Eq.~\eqref{eq:model}, is a sine-Gordon model  with $m_K$ controlling the strength of the non-linearity ($\sim m_K^2\cos \phi$), and the misfit parameter $\delta$ coupling to the gradient of the field ($\sim\delta \partial_x\phi$). This  model is known as  Pokrovsky-Talapov field theory~\cite{AristovLuther,pokrovsky1984theory,PT2d}, and it finds numerous applications in 
	ultra-cold gases~\cite{lovas2022many,KasperMarino,PolkovnikovQBreathers,PhysRevB.75.174511,PhysRevA.68.053609,Buchler_C_IC,haller2010pinning},   trapped ions applied to nano-friction~\cite{Vuletic_nanofriction,liu2019recent,vanossi2020structural},   magnetism~\cite{Bak_1982,Caux_Heisenberg,braun2004frenkel,chepiga2022kosterlitz} and tangentially also  strongly correlated systems~\cite{PhysRevB.66.073105,PhysRev.163.376}. 
	This naturally suggests broad applicability    of our results to  different platforms.

	In addition to the trapped ion simulators discussed above, another example occurs in the cold atoms implementation of Ref.~\cite{KasperMarino}, where the Pokrovsky-Talapov model describes the dynamics of the phase difference between two Raman tunnel-coupled one-dimensional Bose quasi-condensates. 
	The wave-vector difference of the Raman beams plays the role of the solitons chemical potential ($\delta$  {in the model of Eq.~\eqref{eq:model}}).
	Quantum quench dynamics would therefore become accessible through interferometric measurements of the correlation functions of the phase difference~\cite{gring2012relaxation,schweigler2017experimental}. 

	These two examples represent just a few of the several interdisciplinary  
	outreaches
	of our results which establish   one more step towards merging the fields of  topological defects and    dynamical phase transitions in low-dimensional  quantum simulators. 
	In the far future one might imagine nano-scale control of topological defects in real materials by quenching high field optical potentials onto material crystals to controllablly affect material properties.
	Experiments in trapped ion quantum simulators may offer the first steps in such a direction.

	\acknowledgments
	We thank  R.J. Valencia-Tortora and V. Vuleti\'{c} for useful discussions. 
	J. M. acknowledges E. Demler and V. Kasper for previous collaborations on related topics. 
	This project has been supported by the Deutsche Forschungsgemeinschaft (DFG, German Research Foundation) – Project-ID 429529648 – TRR 306 QuCoLiMa (``Quantum Cooperativity of Light and Matter''), and  by the Dynamics and Topology Centre funded by the State of Rhineland Palatinate and Topology Centre funded by the State of Rhineland Palatinate.
	The authors gratefully acknowledge the computing time granted on the supercomputer MOGON 2 at Johannes Gutenberg-University Mainz (hpc.uni-mainz.de).

	\bibliography{EPL_solitons_Short}{} 
	\bibliographystyle{eplbib}
\end{document}